\documentclass[11pt]{article}
\usepackage[utf8]{inputenc}
\usepackage{amsmath}
\usepackage{amssymb}
\usepackage{geometry}
\usepackage{xcolor}
\usepackage{hyperref}
\usepackage[skip=6pt plus 1pt, indent=20pt]{parskip}
\newcommand{\Var}{\text{Var}}
\title{Multiplicative Langevin Process for Volatilities\\
Produces Observed Q-Variance Regularities}
\author{William H. Press\thanks{Corresponding Author: email wpress@utexas.edu}\\
Oden Institute for Computational Engineering and Sciences\\ 
The University of Texas at Austin\\
and\\
Alex Dannenberg\thanks{email: alex@pinemountaincapital.com}\\
Pine Mountain Capital Management\\}
\date{\today}

\begin{document}

\maketitle

\begin{abstract}
Q-variance (so-called) posits a statistical relationship $\mathbf{E}(\sigma^2 | z) = \sigma_0^2 + \tfrac{1}{2}z^2$ between an asset's volatility $\sigma^2$, as observed in a time interval $T$, and its (suitably scaled) return $z$ in the same interval. We here show that this relationship is {\em exactly equivalent} to to positing an Inverse Gamma probability distribution for $\sigma^2$ itself. We then show that such a distribution is exactly generated by a multiplicative Langevin process with an arbitrary, settable coherence time $\tau_c$, so that very nearly the same Q-variance relationship will hold for all $T \ll \tau_c$.
\end{abstract}

\section{Introduction}

Orrell \cite{orrell,wilmott} has pointed out a surprisingly persistent regularity in asset price data, not explainable by a simple random walk model such as the now-long-standard Geometric Brownian motion (GBM) model \cite{black1973}. If $x$ (a random variable) denotes the logarithmic return over a time $T$, then, in a random walk model, the scaled $z=x/\sqrt{T}$ is a random variable that does not, under GBM, depend on $T$.

Now divide intervals of length $T$ into $n$ equal subintervals with respective scaled returns $z_i$, so that $\sum_i z_i = z$. Consider the subinterval variance $\sigma^2 \equiv \Var(z_i)$. A property of the GBM model is that $z$ and $\sigma^2$ are statistically independent (see Appendix A1) .

Notwithstanding, Orrell finds, across a large corpus of asset price data, that $\sigma^2$ and $z$ are not independent, and that, moreover, the conditional expectation $\mathbf{E}(\sigma^2 | z)$, when measured by binning data by $z$ values, empirically follows
\begin{equation}
    \mathbf{E}(\sigma^2 | z) = \sigma_0^2 + k z^2,
\label{qvar}
\end{equation}
Here $\sigma_0$ is a parameter that depends on the specific asset, while $k$ is observed to be close to the value $1/2$. Importantly, the parameters $\sigma_0$ and $k$ for an asset are approximately independent of $T$. Orrell coins the term ``Q-variance" for this phenomenon.

We see three distinct mysteries in the Q-variance regularities: (1) Why the observed universal functional form of equation \eqref{qvar}? (2) Why the observed independence of $T$. (3) Why the value $k=1/2$? 

That observed price movement statistics are fat-tailed (that is, not Gaussian) has long been observed \cite{mandelbrot,fama,heston}. One large class of models explains this as the unfolding of a random walk not in clock time, but according to some latent variable, a ``rate of information arrival". These so-called subordinated time models \cite{clark1973,karpoff1987} make the probability distribution of price changes not Gaussian, but fat-tailed as a mixture of Gaussian distributions. Equivalently, one can regard an asset's volatility $\sigma^2$ not as a constant, but as time varying $\sigma^2(t)$, the result of some stochastic process. Over long times, any such process, if stationary, implies some probability distribution $P(\sigma^2)$.

In the context of such models, this note looks at the implications of the three Q-variance mysteries. After defining some appropriate quantities in \S\ref{sec2}, we proceed in three steps: First (\S\ref{sec3}) we show that the Q-variance functional form of equation \eqref{qvar} uniquely implies that the marginal probability $P(z)$ must be a Student t-distribution. Second (\S\ref{sec4}) we show that this marginal $P(z)$ uniquely implies an Inverse Gamma distribution for the probability $P(\sigma^2)$. Third (\S\ref{sec5}) we show that this Inverse Gamma probability distribution is readily generated by a multiplicative Langevin process whose coherence time is adjustable to be longer than the largest $T$ for which the Q-variance is seen to hold, thus making equation \eqref{qvar} approximately independent of $T$.

\section{Conditional Probabilities and Expectations}
\label{sec2}

We henceforth write $V\equiv \sigma^2$.
The joint distribution $P(V, z)$ is given by
\begin{equation}
P(V, z) = P(z|V)P(V)
\end{equation}
When $z \sim \mathcal{N}(0, V)$, we have
\begin{equation}
P(V, z) = \frac{1}{\sqrt{2\pi V}} \exp\left( -\frac{z^2}{2V} \right)P(V)
\end{equation}
Bayes' rule is
\begin{equation}
P(V|z) = \frac{P(z|V)P(V)}{P(z)}
\end{equation}
where $P(z)$ is the marginal,
\begin{equation}
P(z) = \int_{0}^{\infty} \frac{P(V)}{\sqrt{2\pi V}} \exp\left( -\frac{z^2}{2V} \right) dV
\label{eq6}
\end{equation}
so,
\begin{equation}
P(V|z) = \frac{ V^{-1/2}P(V) \exp\left( -\frac{z^2}{2V} \right) }{ \int_{0}^{\infty} V^{-1/2}P(V) \exp\left( -\frac{z^2}{2V} \right) dV }
\end{equation}

The expectation of the variance $V$ given a constant $z$ is the first moment of the distribution, a ratio of integrals,
\begin{equation}
E[V|z] = \frac{ \int_{0}^{\infty} V^{1/2} P(V) \exp\left( -\frac{z^2}{2V} \right) dV }{ \int_{0}^{\infty} V^{-1/2} P(V) \exp\left( -\frac{z^2}{2V} \right) dV }
\label{eq8}
\end{equation}

\section{Q-variance Regularities Imply Student's t-Distribution for the Marginal $P(z)$}
\label{sec3}

Suppose that $E[V|z]$ is exactly $V_0 + k z^2$ (equation \eqref{qvar}). Then equations \eqref{eq6} and \eqref{eq8} imply
\begin{equation}
\int V \frac{1}{\sqrt{2\pi V}} e^{-z^2/2V} P(V) dV = (V_0 + k z^2) P(z)
\label{eq8a}
\end{equation}
Differentiate both sides with respect to $z$:
\begin{equation}
\int V \left( -\frac{z}{V} \right) P(z|V) P(V) dV = \frac{d}{dz} [(V_0 + k z^2) P(z)]
\end{equation}
The left side is just $-z P(z)$, so,
\begin{equation}
-z P(z) = 2kz P(z) + (V_0 + kz^2) \frac{dP(z)}{dz}
\end{equation}
Rearranging, we have
\begin{equation}
\frac{1}{P(z)} \frac{dP(z)}{dz} = -\frac{(1 + 2k)z}{V_0 + kz^2}
\end{equation}
which can be integrated to give
\begin{equation}
P(z) \propto \left( 1 + \frac{k}{V_0} z^2 \right)^{-\left( \frac{1}{2k} + 1 \right)}
\label{eq13}
\end{equation}
which is Student's t-distribution with parameter $\nu = 1/k + 1$.

Equations \eqref{eq8a}--\eqref{eq13} can be read in reverse order to prove the converse, that a Student $t$ distribution for $P(z)$ implies the Q-variance regularity for $E[V|z]$.

\section{Student's $t$ Marginal Uniquely Implies Inverse Gamma Conditional }
\label{sec4}
Here, we show that the Inverse Gamma distribution is the the unique mixing distribution that produces Student’s t-distribution from a Gaussian scale mixture, hence is the unique distribution implied by the observed Q-variance functional form.

We seek to solve for $P(V)$ using the integral equation defined by the Gaussian scale mixture and the empirical marginal distribution (equations \eqref{eq6} and \eqref{eq13}),
\begin{equation}
\int_{0}^{\infty} \frac{1}{\sqrt{2\pi V}} \exp\left( -\frac{z^2}{2V} \right) P(V) dV \propto \left( 1 + \frac{k}{V_0} z^2 \right)^{-\left( \frac{1}{2k} + 1 \right)}
\end{equation}

First, we change variables to turn the Gaussian kernel into a standard Laplace Transform. Let $s = z^2/2$ and $w = 1/V$. The differential becomes $dV = -(1/w^2) dw$. The limits $V \in (0, \infty)$ map to $w \in (\infty, 0)$. Substituting these into the integral, we obtain:
\begin{equation}
\int_{0}^{\infty} e^{-sw} \left[ w^{-3/2} P(1/w) \right] dw \propto \left( 1 + \frac{2k}{V_0} s \right)^{-\left( \frac{1}{2k} + 1 \right)}
\label{eq15}
\end{equation}
The left side is exactly the Laplace Transform of the function $f(w) = w^{-3/2} P(1/w)$. 

The inverse Laplace Transform of the right-hand side can be found using the standard table result,
\begin{equation}
\mathcal{L} \left\{ \frac{\beta^\alpha}{\Gamma(\alpha)} w^{\alpha-1} e^{-\beta w} \right\} = \left( 1 + \frac{s}{\beta} \right)^{-\alpha}
\end{equation}
yielding
\begin{equation}
\mathcal{L}^{-1} \left[\left( 1 + \frac{2k}{V_0} s \right)^{-\left( \frac{1}{2k} + 1 \right)}\right]
 \propto w^{\frac{1}{2k}} \exp\left( -\frac{V_0}{2k} w \right)   
\end{equation}
Equating the Laplace transforms of the two sides give
\begin{equation}
w^{-3/2} P(1/w) \propto w^{\frac{1}{2k}} \exp\left( -\frac{V_0}{2k} w \right)
\end{equation}
and replacing $w$ with $1/V$,
\begin{equation}
P(V) \propto V^{-\left( \frac{1}{2k} + \frac{3}{2} \right)} \exp\left( -\frac{V_0}{2k V} \right)
\end{equation}
This shows that $P(V)$ is uniquely an Inverse Gamma distribution with parameters
\begin{equation}
    \alpha = \frac{1}{2k} + \frac{1}{2} = \frac{1}{2}\nu, \qquad \beta = \frac{V_0}{2k}
\end{equation}

While this calculation does not indicate that the claimed value $k\approx 1/2$ should be special, it allows that value to be obtained by choosing the Inverse Gamma parameter $\alpha = 3/2$. 

\section{Stochastic Process Model for $\sigma^2(t)$ }
\label{sec5}
Thus far, we have ``explained" the Q-variance functional form as deriving uniquely from an Inverse Gamma distribution for the probability distribution $P(\sigma^2)$. The value of the coefficient $k\approx 1/2$ is left as parameter deriving somehow from the psychology of markets, to be fitted from the data.

The remaining mystery, the invariance of Q-variance over a wide range of segment sizes $T$ can be simply viewed as a statement that the stochastic process $\sigma^2(t)$ should have a coherency time $\gg T_\text{max}$, the maximum $T$ studied.

The challenge, then, is to exhibit a stochastic process whose stationary solution is Inverse Gamma distributed, but that also has an adjustable coherency time. That is straightforward. The Langevin process with multiplicative noise, \cite{huang2,huang,sornette}
\begin{equation}
dV_t = \gamma (\bar{V} - V_t) dt + s V_t dW_t
\label{langevin}
\end{equation}
where $\gamma$ is the mean-reversion rate, $\bar{V}$ is the target level, and $s$ is the noise strength, is easily shown (see Appendix A2) to produce 
the stationary solution to the Fokker-Planck equation,
\begin{equation}
P(V) \propto V^{-(2 + 2\gamma/s^2)} \exp\left( -\frac{2\gamma\bar{V}}{s^2 V} \right)
\end{equation}
with a coherency time (see Appendix A3)
\begin{equation}
\tau_c \approx \frac{1}{\gamma + \tfrac{1}{2}s^2}
\label{eq22}
\end{equation}
Mapping these to the Inverse Gamma parameters $\alpha$ and $\beta$ gives
\begin{equation}
\alpha = 1 + \frac{2\gamma}{s^2}, \quad \beta = \frac{2\gamma\bar{V}}{s^2}
\end{equation}
Alternatively, one can specify any desired values, for example, $k = \frac{1}{2}$ and $V_0\equiv \sigma^2_0$, and any sufficiently large value $\tau_c$, and then compute
\begin{equation}
    \gamma = \frac{1-k}{(1+k)\tau_c},\qquad
    s^2 = \frac{4k}{(1+k)\tau_c},\qquad
    \bar{V}=\frac{V_0}{1-k}
\label{eq24}
\end{equation}
The resulting multiplicative Langevin process equation \eqref{langevin} will then ``explain" all three Q-variance mysteries:
(1) $\mathbf{E}(\sigma^2 | z) = \sigma_0^2 + k z^2$, (2) with $\sigma_0^2, k$ as specified, and (3) invariant with $T$ for $T \ll \tau_c$.

Positing a multiplicative Langevin process introduces a new parameter that can be taken as either $\gamma$ or else $s^2$ separately, or else as the coherence time $\tau_c$ (equation \eqref{eq22}), and cannot be estimated from returns as an unordered set.
As $T$ approaches or exceeds $\tau_c$, mean reversion will begin to pull the variance toward $\bar{V}$, causing $\mathbf{E}(\sigma^2 | z)$ to plateau and driving $k$ in equation \eqref{qvar} to zero. Any Gaussian mixture model thus implies that the Q-variance regularity cannot persist for all $T$, but must disappear as $T$ gets large.

\section{Relation to Previous Work}

Stochastic models for time-varying variances, in both continuous and discrete time, have long been established in the finance literature \cite{hull1988, scott1987, wiggins1987}. Some widely used models, however, do not naturally yield the fat tails required to explain observed price statistics. For instance, the Heston model \cite{heston} employs an affine square-root process resulting in a Gamma stationary distribution for the variance, which lacks the heavy tails seen in empirical data. Other early models, such as Hull and White \cite{hull1988}, did not incorporate mean reversion.  

Nelson \cite{nelson1990} notably demonstrated that the discrete-time GARCH(1,1) model converges to the multiplicative Langevin process (equation \eqref{langevin}) in the continuous limit. This process, often called Nelson's diffusion, possesses both desirable properties: it is mean-reverting and yields a stationary Inverse Gamma distribution with power-law tails. Barone-Adesi et al. \cite{barone2004} have explored the implications of this framework for option pricing.  

The contribution of this paper is to situate the ``Q-variance" regularity observed by Orrell \cite{orrell} within this established framework, providing a unified derivation that explains its functional form, its invariance over time scales $T$, and its characteristic coefficient.

\section{Summary}
The assertion (``Q-variance") that $\mathbf{E}(\sigma^2 | z) = \sigma_0^2 + k z^2$ holds with $k=1/2$ is exactly equivalent to the statement that $P(\sigma^2)$ is Inverse Gamma distributed with $\alpha = 3/2$. Any other distribution $P(\sigma^2)$ will produce a different $\mathbf{E}(\sigma^2 | z)$. Whether this value of $k$ or $\alpha$ is universal, or is supported by any particular data set, is an issue to be discussed elsewhere.

The required $P(\sigma^2)$ is readily produced by a multiplicative Langevin process (also known as Nelson's diffusion process) whose time coherence $\tau_c$ can be chosen to be much longer than any measured subinterval size $T$, producing arbitrarily good $T$ independence for that and any smaller value of $T$.

Apart from $\sigma_0^2$, which is observed to be different for different assets, the model has only one adjustable parameter, $\alpha$ or $k$, and arguably zero arbitrary choices of functions (unless one counts Langevin, equation \eqref{langevin}, as arbitrary).

\section*{Disclosure of Use of AI}
Gemini (``3 Fast") and ChatGPT (GPT-5.3) assisted in developing and checking the equations in this paper. Each LLM checked the other's version of the math, finding no significant discrepancies. We provided overall guidance and a final human check. The LLMs were not used in writing the manuscript. We are responsible for all errors in both the math and the words.  

\section*{Appendices}
\subsection*{A1. Random Walk Implies That $z$ and $\sigma^2$ Are Statistically Independent}

Let $\mathbf{1}$ be vector $(1,1,1,\ldots,1)$ of length $n$. Let $\mathbf{Z} = (z_1,z_2,\ldots,z_n) \sim \mathcal{N}(0,\sigma^2\mathbf{I})$ be the vector of scaled returns in the subintervals. Now decompose $Z$ into vectors along and perpendicular to $\mathbf{1}$:
\begin{align*}
    Z_\parallel &= (\bar{z}, \bar{z},\ldots, \bar{z})\\
    Z_\perp &= (z_1-\bar{z}, z_2-\bar{z},\ldots, z_n-\bar{z})
\end{align*}
where $\bar{z} \equiv (1/n)\sum z_i$. Since these are orthogonal directions in the spherically symmetric normal distribution $\mathcal{N}(0,\sigma^2\mathbf{1})$, they are statistically independent. Since the total return $z = Z_\parallel \cdot \mathbf{1}$ is a function only of $Z_\parallel$, while $\sigma^2 = \frac{1}{n-1}Z_\perp\cdot Z_\perp$ is a function only of the second, $z$ and $\sigma^2$ must also be statistically independent.

\subsection*{A2. Multiplicative Langevin Process Yields Inverse Gamma}
To get the Fokker-Planck stationary solution for the process
\begin{equation}
dV_t = \gamma(\bar{V} - V_t)dt + s V_t dW_t
\label{lang2}
\end{equation}
we equate the drift and diffusion terms, so that there is no net flux of probability:
\begin{equation}
\gamma(\bar{V} - V)P(V) = \frac{1}{2} \frac{d}{dV} \left[ s^2 V^2 P(V) \right]
\end{equation}
Defining $G(V) = s^2 V^2 P(V)$, this becomes
\begin{equation}
\frac{\gamma(\bar{V} - V)}{s^2 V^2} G(V) = \frac{1}{2} \frac{dG}{dV}
\end{equation}
whose solution is
\begin{equation}
G(V) \propto V^{-2\gamma/s^2} \exp\left( -\frac{2\gamma \bar{V}}{s^2 V} \right)
\end{equation}
or
\begin{equation}
P(V) = \frac{G(V)}{s^2 V^2} \propto V^{-\left( \frac{2\gamma}{s^2} + 2 \right)} \exp\left( -\frac{2\gamma \bar{V}}{s^2 V} \right)
\end{equation}
which is an Inverse Gamma distribution with
\begin{equation}
\alpha = \frac{2\gamma}{s^2} + 1, \qquad \beta = \frac{2\gamma \bar{V}}{s^2}
\end{equation}

\subsection*{A3. Calculation of Coherency Time}
We transform equation \eqref{lang2} into an Ornstein-Uhlenbeck (OU) process by changing variables to $x=\log V$ via Itô's Lemma. This yields
\begin{equation}
dx = \left[ \gamma \bar{V} e^{-x} - \left(\gamma + \frac{1}{2}s^2\right) \right] dt + s\,dW_t
\equiv A(x)dt + s\, dW_t
\end{equation}
The equilibrium is found by setting to zero the drift term, $A(x)=0$, implying immediately
\begin{equation}
    \gamma \bar{V} e^{-x} = \gamma + \frac{1}{2}s^2 
\end{equation}
Near equilibrium, we can linearize the equations, in which case
the relaxation rate $\lambda$ (reciprocal of relaxation time $\tau_c$) becomes the negative derivative of $A(x)$ at equilibrium, 
\begin{equation}
\lambda = -(-\gamma \bar{V} e^{-x}) = \gamma + \frac{1}{2}s^2
\end{equation}

\bibliographystyle{unsrturl}
\bibliography{sample}

\end{document}